# Analyzing Answers in Threaded Discussions using a Role-Based Information Network

Jeon-Hyung Kang and Jihie Kim

University of Southern California / Information Sciences Institute
4676 Admiralty Way, Marina del Rey, CA 90292, USA
{jeonhyuk, jihie}@isi.edu

*Abstract*—Online discussion boards are an important medium for collaboration. The goal of our work is to understand how messages and individual discussants contribute to Q&A discussions. We present a novel network model for capturing information roles of messages and discussants, and show how we identify useful answers to the initial question. We first classify information seeking or information providing roles of messages, such as question, answer or acknowledgement. We also identify user intent in the discussion as an information seeker or a provider. We capture such role information within a reply-to discussion network, and identify messages that answer seeker questions and how answeres are acknowledged. Message influences are analyzed using B-centrality measures. User influences across different threads are combined with message influences. We use the combined score in identifying the most useful answer in the thread. The resulting ranks correlate with human provided ranks with an MRR score of 0.67.

## I. INTRODUCTION

Online discussion boards play an important role in various fields, including science, politics, and education. In web-enhanced courses, discussion boards are heavily used for question answering and collaborative problem solving ([8], [10]).

Discussion threads represent conversational dialogue among the participants where each discussion thread consists of a set of messages organized according to reply-to relations or temporal orderings. In Q&A (Questions and Answers) discussions, one of the interesting challenges is identifying the most useful or influential answer to the initial question. Such messages can help others who have similar questions. It also helps assessing contributions by individual participants.

For modeling answer strengths, we keep track of how information flows in discussions. The Forum Speech Act represents roles of individual message such as question, answer, or acknowledgement [8]. Such Q&A roles of individual messages indicate how information flows from the answers to the corresponding questions in the given discussion thread.

Since each discussion thread focuses on specific questions that are closely related, the role of a user as an information seeker or an information provider usually remains the same within the thread. This is especially the case in discussions over a limited period of time. Such thread-level user role information can give additional hints on the roles that the messages posted by the user play within the thread. For example, information provider's questions tend to request additional details of the given problem rather than raising additional issues or seeking help.

Finally, the global information influence of individual users such as the degree of useful information provided by a user in the whole forum can give hints on the usefulness of the message posted by the user.

In this paper, we present a novel framework called RiNet (Role-based Information Network) for generating an information flow network from online discussion threads. Each node represents a user or a message. Edges are generated from authorship and message reply-to relations. Using the user role and message role information, we identify channels where answers are provided to information seekers and how they are acknowledge or challenged. The information network connects the messages posted by the same user in different threads through the authorship and represents the user's global roles.

Once the RiNet for all the threads in the given forum is built, the overall influence of each node is computed using centrality measures. For computing message influence scores, in order to incorporate the author's overall influence in the whole forum, we accumulate the author influence scores as well as the direct influence scores of the message within the thread. In finding useful answers, messages are ranked according to these scores. Currently, the resulting ranks correlate with the rankings provided by human annotators with a Mean Reciprocal Rank (MRR) score of 0.67.

The paper first presents our framework for modeling and classifying messages and participants with respect to their information roles (Section II, IV). We then show how RiNet is created from the role information. (Section III). The resulting model is used in identifying the most useful response to the initial question (Section V). Finally, we summarize the current results and discuss future work. The main contributions of the paper are:

1. Presenting a novel RiNet framework that captures rich semantics of (a) message information roles, (b) thread-level user roles, and (3) global user roles within the Q&A forum.

2. Providing an approach for identifying the most useful response to the initial question in student Q&A discussions.

Our work takes place in the context of an undergraduate course discussion board that is an integral component of an Operating Systems course in the Computer Science department at the University of Southern California. The course is offered every semester, and always taught by the same instructor. The students use the discussion board, most commonly, to seek help on project assignments. For this study, we use data from the Spring 2006 semester and the Fall 2007 semester.

## II. INFORMATION ROLES OF MESSAGES AND USERS

As a step toward identifying useful responses, we first identify information roles of messages and participants. Analyzing individual messages with respect to their true information seeking or information providing roles is challenging. First of all, we need to handle noises and the informal nature of student discussion text. Second, standard surface-level grammatical forms are not enough in distinguishing questions from answers and vice versa. Answers are commonly provided in a form of a question, such as "Have you checked the Nachos Manual section 4.3?" Surface-level features such as 'wh' words (what, where, when, how) or punctuations, such as a question mark, are not sufficient. Moreover, some questions are posted in order to provide help rather than seeking help, as in the case of a request to elaborate the given question. For example, "What do you mean by X?" can play either an information seeking or an information-providing role depending on the context. If we apply grammatical analysis or typical speech act analysis on discussion threads without considering author and message roles, it is difficult to capture such differences.

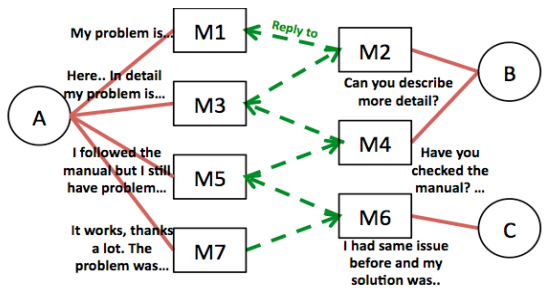

Figure 1. An example discussion thread

Figure 1 shows an example discussion threads with a sequence of seven messages: M1, M2, ..., M7 in order. User A, ..., C represent the discussants. User A initiates the thread by describing the problem that he or she has (M1). Questions don't necessarily take a grammatical form of a question, as in this case. In message (M2), user B asks for more details about user A's problem. User A then sends problem details (M3). B provides help by sending a pointer (M4), etc. Note that the answer in M4 is given in a form of question. Furthermore, although M3 is an answer to M2, the role of M3 as the information source (answer) to the initial question should not be as significant as the answer role of M4 since it only elaborates the initial question. As a user level analysis, different users are participated in the discussion board with different intention. User A participated in this thread to ask for help as an information *seeker* while user B and C are participated in this discussion to help others as an information *provider*.

If a message asks for information, we call it as a *sink* message (e.g. M1, M2, and M5 in Figure 1), while if one provides information, a solution, or an answer; we call it as a *source* message (e.g. M3, M4, and M6 in Figure 1). Significance of a message can be determined based on both the message role and the poster role. That is, influence of M3 as a source can be reduced since it is posted with the intention of seeking help. Therefore, a thread-level context analysis that considers **participant roles and intentions** are needed. Other dialogue information such as a positive acknowledgement for an answer (e.g. "Thank you".) and patterns of user-to-user interactions with different types of messages (e.g. seeker-sink-source-provider) can give additional hints on the significance of the answer.

## III. RINET: ROLE-BASED INFORMATION NETWORK

This section describes how we generate information flow network from discussion data.

### A. Information roles of adjacent messages

In order to understand the nature of information flow between adjacent messages, we manually examined 248 randomly picked message pairs that are connected by 'reply-to' relations. An edge is either a reply-to relation between two messages or an authorship of a message by a user, as shown in Figure 2. It also lists the most frequent patterns.

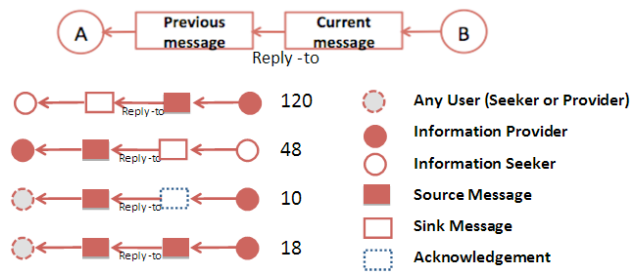

Figure 2. Modeling messages connected by a 'reply-to'

We model each pair (user-message-message-user) with respect to information roles (sink/source/seeker/provider) and positive/negative acknowledgements. Figure 2 shows four most popular pairs with their frequency of occurrence. The most frequent pair is *Seeker-Sink-Source-Provider* representing a seeker question followed by a provider answer. We call such paths as *answer* channels.

On the other hand, *Provider-Sink-Source-Seeker* indicates elaboration of the prior question as described above. We call them as *elaboration* channels. Their sources don't directly provide answers to the seeker question.

In general, a positive/negative acknowledgement of an answer expresses satisfaction/dissatisfaction by the answer. We represent such a pair (a source followed by an acknowledgement) as an *ack* channel. Note that acknowledgement does not provide any information as either a source or a sink message, but it just indicates usefulness of the information.

We also examined how neighbors of sources provide additional hints on usefulness of the answer. *Provider-Source-Sink-Seeker* occurs when given answer was not satisfied (i.e. does not work, unclear, how question, confirmation question) or prompted questioner to next related problem (i.e. it worked but now I have another problem). That is, when a Sink follows a Source, it often indicates dissatisfaction.

*Provider-Source-Source-Provider* occurs when another answer from a provider corrects the given answer. On the other hand, *Provider-Source-Source-Seeker* occurs when Seekers find their own solutions and share them with others. Overall, when a source is followed by another source, it often implies weakness of the prior source.

Based on these observations, we included additional *ack* channels for cases when either a sink or another source follows a source.

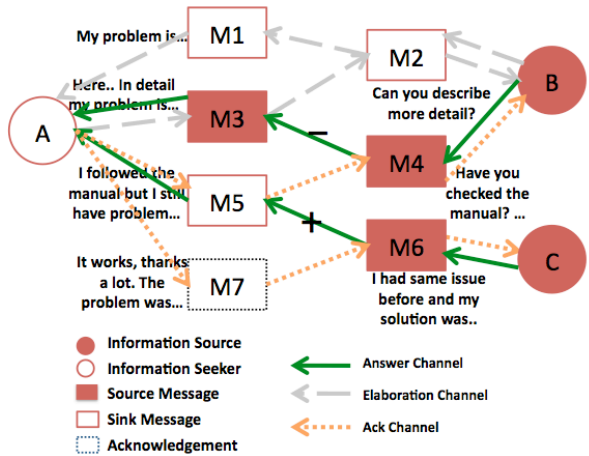

Figure 3. An example RiNet with different Information Channels

### B. Information network generated from Message and Poster Roles

We represent three channels: *answer*, *elaboration* and *ack*, as described above. We first start with the reply-to links with authorship, as shown in Figure 2. We then mark information roles of individual messages and authors. We then assign channel types to the reply-to edges. Figure 3 shows the RiNet model of the discussion in Figure 1. M4->M1 and M6->M5 represent answer channels. M2->M1 and M3->M2 are elaboration channels. Also, M7->M6 is a positive ack channel, and M5->M4 is a negative ack channel.

Given these channels, we can assign different edge weights depending on how we want to model the information flow. For example, weights for edges on elaboration channels can be reduced or minimized, as they don't directly provide an answer to the initial question. The positive/negative ack channels can increase/decrease weights of the prior sources as they strengthen/weaken the answer. We can control the effect of the ack channel in increasing/decreasing weights of the corresponding source with a parameter $\gamma$.

## IV. GENERATING CLASSIFIERS FOR MESSAGE AND PARTICIPANT ROLES

### A. Data Annotation

We adopted a Speech Act framework for Q&A dialogue ([14], [8]) and extended it to capture the true information roles of the messages and the users. Individual messages were annotated with sink and source information. The same message can have both sink and source roles with respect to the prior message or its poster. Human annotators were also asked to tag the region of the messages that helped them identify sources and sinks. In order to overcome data sparseness, word feature generation focuses on those regions in the training data. For user roles within a thread, the annotator marked the role of individual participants as an information provider and an information seeker. The annotation scheme was developed over three years by multiple (> 6) annotators until they reach enough agreement. The annotators shared and compared their annotations while they were developing the annotation scheme. For inter-annotator agreement, the kappa values were computed with an independent dataset that consists of 30 discussion threads with 99 messages to measure of agreement between the two individuals. Kappa is a robust measure for agreement that takes into account agreement by chance [21]. 1 implies perfect agreement while 0 means no agreement among the annotators other than by chance. Two annotators participated. For all the categories, the annotators show a high level of agreement. Negative acknowledgements were excluded as only 3% of the data has them. Kappa score of sink, source message, information seeker/provider, and positive acknowledgement is 92.92, 95.95, 98.98 and 87.00% respectively.

### B. Data Processing and Feature Generation

Because student discussions are very informal and noisy with respect to grammar, syntax and punctuation, our model fixes common typos, transforms informal words to formal words, and converts apostrophes to their original forms. For example, "I'm" is converted to "I am". Informal words are also substituted by formal words. For example, "ya", "yea", and "yup" are all substituted by "yes". We leave question marks, but other punctuations such as brackets, colons, comma, dashes, and ellipses are all replaced with a blank space, as they were not helpful for identifying message or user roles. We also perform word stemming.

To reduce the variance, we generated 20 word categories from a randomly selected training set: "I" and "we" were replaced by CAT_SUB_I_WE, "is", "was", "are", "were" were replaced by CAT_BE, interrogative words such as "what", "where" were replaced by CAT_WH and "fault", "problem", and "error" were categorized by CAT_ISSUE. We then created unigram, bigram, and trigram combinations of the processed words. After the text-preprocessing step we could decrease the size of unigram vector size from 5257 to 457, bigram vector size from 3025 to 1218 and trigram vector size from 1041 to 1031. To classify source and sink, in addition to the standard n-grams features, we included their positions (beginning, middle, or end position). The positions of the cue words are important because those one in beginning sentences can have different meanings than those in subsequent sentences. For example, "Thank you" in the beginning sentence position may be an expression of gratitude for previous information, while "thank you" in the last sentence may indicate only politeness. We also used thread-level features, including author change information, the relative position of the message in the thread, the author's participation frequency (normalized), sequence number of message, n-grams of previous messages with their positions, and previous author information such as participation frequency. Table 1 shows top 5 high frequency unigrams, bigrams, and trigrams list for source messages.

The 2006 spring and 2007 fall semester discussion threads were randomly divided 240 threads (904 messages) into two datasets: 180 threads (634 messages) for training and 60 discussion threads (270 messages) for testing.

We used a Support Vector Classifier [19] and Random Forest [20] in the WEKA package to create two binary classifiers that identify message roles: source and sink and one classifier for user roles: information seeker and information provider. A message can have both source and sink role with respect to the prior message, or neither of them (e.g. acknowledgement). However, a user tends to play either a seeker or a provider role in a given thread.

TABLE I. TOP 5 UNIGRAMS, BIGRAMS, AND TRIGRAMS

| Unigram | Bigram | Trigram |
|---|---|---|
| CAT_SUBJECTIVE_IWE | CAT_SUBJECTIVE_IWE CAT_BE | DO CAT_SUBJECTIVE_IWE NEED |
| CAT_BE | CAT_SUBJECTIVE_IWE HAVE | CAT_SUBJECTIVE_IWE CAT_BE GET |
| ? | DO CAT_SUBJECTIVE_IWE | CAT_SUBJECTIVE_IWE CAT_BE NOT |
| CAT_U | PART CAT_NUM | CAT_WH DO CAT_SUBJECTIVE_IWE |
| CAT_WH | CAN CAT_SUBJECTIVE_IWE | CAT_SUBJECTIVE_IWE DO NOT |

TABLE II. TEST SET RESULT

| Model | Classifier | Prec. | Recall | F-Score |
|---|---|---|---|---|
| Support Vector | Sink message | 0.88 | 0.88 | 0.88 |
| | Source message | 0.83 | 0.80 | 0.83 |
| | Seeker/Provider | 0.84 | 0.84 | 0.84 |
| | Positive Ack | 0.74 | 0.36 | 0.48 |
| Random Forest | Sink message | 0.88 | 0.88 | 0.88 |
| | Source message | 0.79 | 0.78 | 0.78 |
| | Seeker/Provider | 0.85 | 0.85 | 0.85 |
| | Positive Ack | 0.64 | 0.534 | 0.58 |

The results of 10-fold cross validations the F-score reaches 0.91 for sink message, 0.88 for source message, 0.86 for info seeker/info provider, and 0.54 for Positive Acknowledgement with the Support Vector classifiers. Table 2 shows our model accuracy compared to our annotated target value. The frequency of Positive Acknowledgement is much lower than others, which affects the classifier accuracies.

## V. PROFILING INFLUENCE OF USERS & MESSAGES WITH CENTRALITY MEASURES

Social scientists often use network models in representing individuals and organizations tied by some relations. Nodes are users or events. Edges represent friendship, authorship, or relationships of beliefs, etc. They use various metrics such as betweenness, bridge, centrality, degree, and eigenvector centrality in analyzing the role of a node. Centrality measures allow us to identify the relative importance of a node with respect to its influence.

Message influences and their aggregation over the network were analyzed using Bonacich centrality measures ([2], [6]). B-centrality measure uses not only the centrality (relative importance) of a node based on given direct and indirect edge connections but also its neighbor nodes position in the global networks. It is calculated by the total number of attenuated paths among nodes with direct links attenuation factor $\beta$ and indirect links attenuation factor $\alpha$. Attenuated paths from node A to node B can be regarded as influence of node A on node B. In other words, if one user participates more discussions with many source messages as an Information Provider, sum of his attenuated paths or his influence on others will be higher than others. By applying b-centrality for discussion networks, we can find who influence a specific node the most or who/what is the most influential node.

The influence matrix C is

$$C = \beta A + \beta \alpha A^2 + \beta \alpha^2 A^3 + ... + \beta \alpha^{(n-1)} A^n$$

where n is the maximum possible message passing length in the forum and A is an influence matrix generated by the thread model described above in Section III. That is, A represents nodes and edges in the RiNet, where entry $i,j$ is not 0 if and only if there is an edge from node $i$ to node $j$, and 0 otherwise.

$$A = \begin{bmatrix} UU_{u \times u} & UM_{u \times m} \\ MU_{m \times u} & MM_{m \times m} \end{bmatrix}$$

Each $(i, j)$ entry in sub-matrix MM represents whether message $i$ replies to message $j$. In sub-matrices UM and MU, the value for user $i$ to message $j$ (or the value for message $i$ to user $j$) is set based on authorship. All the diagonal values and UU values are zero.

For example, the two-hop (via two edges) influence from node $i$ to node $j$ is given by $\beta * \alpha * A^2$ in its $(i, j)$ entry. The global influences from node $i$ to node $j$ is given by $C(i, j)$. Note that when we consider globally influential or important nodes, we need to use a large alpha value since a large alpha value raises the effect of long distance interactions. We used 9 as our n value, since a sequence of 3 people can connect two different discussion threads via the person who belongs to both, and 3 people can be reached with 9 hops. Different n value can be used based on the characteristics of discussion forum. For beta (direct attenuation), we are currently using 1 and for alpha we are using 0.1 to focus on other parameters.

To find the most important answers for the given question, we need to combine author's normalized global source influence and message influence on the question. Total global influence of node $i$ that reached other nodes are measured by

$$|C_i(\alpha,\beta)| = \sum_j C_{ij}(\alpha,\beta)$$

and normalized global source influence of node $i$ is measured by $|\overline{C}_i(\alpha,\beta)| = |C_i(\alpha,\beta)|/N_i$ where $N_i$ is total number of edges coming in/out from node $i$. The most important answer for sink message j can be

$$S_i = \underset{\{(Thread(i)=Thread(j)) \cap Source(i)\}}{\arg\max} ((1-\omega) \times |C_i(\alpha,\beta)| + \omega \times |\overline{C}_{Author(i)}(\alpha,\beta)|)$$

where $|C_i(\alpha,\beta)|$ is source node $i$'s source influence in the thread and $|\overline{C}_{Author(i)}(\alpha,\beta)|$ is normalized global influence of source node $i$'s author. Note that we only consider candidates that belong to the same thread and scores with respect to sink node j, the initial question. With a coefficient $\omega$, we combine the influence of the author. When $\omega$ is 0, we are using message influences only, while when $\omega$ is 1, we are considering global factor only. The resulting score is used for

TABLE III. MRR SCORES FOR DIFFERENT INFLUENCE MODELS USING INFORMATION ROLE CLASSIFIERS

| Ranking Strategy | Information used | MRR |
|---|---|---|
| *1.1 Influence Network Model score with $\gamma$ =1.0 & $\omega$ =1.0* | User role & Message role & Positive Ack | 0.63 |
| *1.2 Influence Network Model score with $\gamma$ =0.5 & $\omega$ =1.0* | | 0.65 |
| *1.3 Influence Network Model score with $\gamma$ =0.5 & $\omega$ =0.5* | | **0.67** |
| *1.4 Influence Network Model score with $\gamma$ =0.5 & $\omega$ =0.0* | | 0.62 |
| *1.5 Influence Network Model score with $\gamma$ =0.0 & $\omega$ =0.0* | | 0.61 |
| *2. Earlier messages from information providers* | User role & Message location | 0.62 |
| *3. Earlier source messages* | Message role & Message location | 0.61 |
| *4. Second message* | Message location | 0.61 |
| *5. Later source messages* | Message role & Message location | 0.53 |
| *6. Later messages from information providers* | User role & Message location | 0.51 |
| *7. Last message* | Message location | 0.44 |

TABLE IV. MRR SCORES FOR DIFFERENT INFLUENCE MODELS USING INFORMATION ROLE FROM HUMAN ANNOTATIONS

| Ranking Strategy | Information used | MRR |
|---|---|---|
| *1.1 Influence Network Model score with $\gamma$ =1.0 & $\omega$ =1.0* | User role & Message role & Positive Ack | 0.75 |
| *1.2 Influence Network Model score with $\gamma$ =0.5 & $\omega$ =1.0* | | 0.82 |
| *1.3 Influence Network Model score with $\gamma$ =0.5 & $\omega$ =0.5* | | **0.85** |
| *1.4 Influence Network Model score with $\gamma$ =0.5 & $\omega$ =0.0* | | 0.78 |
| *1.5 Influence Network Model score with $\gamma$ =0.0 & $\omega$ =0.0* | | 0.72 |
| *2. Earlier source messages* | Message role & Message location | 0.70 |
| *3. Earlier messages from information providers* | User role & Message location | 0.68 |
| *4. Later source messages* | Message role & Message location | 0.55 |
| *5. Later messages from information providers* | User role & Message location | 0.55 |

selecting the most influential messages.

If a message has a higher source score than others, it is considered as the most influential and informative response to the given question based on message-to-message influence and author influence. Likewise, if a message has higher sink score than other messages, it can be considered as the more influential question (or more interesting problems) compared to others, since it generated many following up discussions.

VI. ANALYZING ANSWER USEFULNESS WITH RINET

To evaluate our source score accuracy, we annotated the most influential source message for the initial question (sink) in each thread. Out of 60 discussion threads in test set, we considered 20 threads as our evaluation set that contained more than three messages. Since shorter threads usually contain only one answer and finding most influential source message for the first sink message is just the same as finding any source message. Selecting the most important source message can be tricky when there are two or three redundant related answers. Depending on the annotator's interpretation, there can be more than one important source messages. Before the evaluation, we checked agreement between two annotators with 25 threads. The kappa value was 0.82 and the F score was 0.65.

We used Mean Reciprocal Rank Score (MRR) in evaluating the scores from the model against human annotations. MRR is a measure for lists of ranked model predictions to a query. It considers rank position of the most relevant message to the question. The score is the average of the inverse of the rank of the correct answer.

$$MRR = \frac{1}{|Q|} \sum_{i=1}^{Q} \frac{1}{rank_i}$$

Q is the number of threads and rank$_i$ is the ranking of correct response messages in the $i^{th}$ thread using the model score. Our model correctly picked the most important answers in 12 threads (row1.3 in Table 4) out of 20 threads when alpha was larger than zero.

To compare different influence models, we computed rankings based on these in addition:

2) *Earlier source messages*: Rank source messages based on closeness to the initial messages.
3) *Earlier message from providers*: Rank messages posted by information providers based on closeness to the initial messages.
4) *Second messages*: Rank source messages based on closeness to the initial messages.
5) *Later source messages*: Rank source messages based on closeness to the last message.
6) *Later messages from providers*: Rank messages posted by information providers based on closeness to the last message.
7) *Last messages*: Rank based on closeness to the last message.

The messages posted by information providers and the ones that are closer to the initial questions can be important. The last messages posted by information providers can be also influential since it may conclude the discussion. Similarly, the source messages that are closer to the initial question or the last source messages can be good candidates. As a simple baseline, important messages can also be identified by their positions in a thread: the message that follows the initial question (i.e. the second message) or the last message in the thread. Note that, in Table 3, strategies 1, 2, 3, 5, and 6 are using message and user role information and 4 and 7 are not using message and user role classifier results.

As expected, our model produces better MRRs. We applied different parameters $\omega$ and $\gamma$ setting and the best MRR can be reached using balanced weights between global user role and local message role (0.67 for 0.5 and 0.5). We expect further improvement by optimizing it over the parameter space. Among the rest, earlier messages from information providers or earlier source messages provide better MRR scores (0.62 and 0.61) than other alternatives. Note that without using any user and message role information, the second message provides the best MRR score (0.61). In order to assess the effect of the classifier accuracies, we computed MRR scores using human annotated role information and compared them with the results with automatically classified role information. As shown in Table 4, similar patterns are observed.

## VII. RELATED WORK

There has been increasing interest in modeling online dialogue including email exchanges, chats, blogs, etc. For example, Talk-to-me [7] can predict the likelihood that a message will receive a reply based on the content of the message and the message sender. Requests and commitments of email exchange are analyzed in [9]. Carvalho and Cohen [3] presents a dependency-network based collective classification method to classify email speech acts. Similar approaches for classifying speech acts were investigated to capture discussion focus [5]. Our work expands these message role analyses by capturing user roles for the given discussion or context, and supports a combination with global user contribution. [4] presents a method for analyzing user communication roles in discussion forums based on 9 features: popularity, reciprocity, and length of interaction, initialization, neighbor's roles and volume of communication measures. However their features do not directly capture message information flow or user intention. Our work complements it by modeling best answers in discussion network.

Chen et al., [22] classifies user reputation in discussion forums based on quality of contributed text. Weimer et al. [23] analyzes quality of forum posts, and reports post relations, such as referencing and reply-to relations as important features. We expect that content quality and additional hints on user reputation will be useful in identifying best sources.

There have been significant advances in understanding networks using community detection algorithms. For example, [11] applies clustering to identify consensus and consensus facilitators. Such information can be useful for us in filtering out discussion threads that don't reach a conclusion, since such discussions may not have useful information sources.

There have been approaches for assessing user participation patterns in forums. [17] utilizes social influence for increasing user participation. Their assumption is users are more motivated to take part in discussion when observing other users' participation. [18] argues that either her friends or her own interests in dynamic multi-topic discussions in online forums motivate user's participation in topic discussion. [15] models information access behavior of a group as an information flow issue. The resulting model is used in recommendation services. [1] studies two different interaction roles among peers and between students and instructors using statistical text analysis. [10] performs qualitative analysis of student reasoning in discussions. We expect that these forum behavior models will be useful in refining the user role and global influence models.

## VIII. SUMMARY AND DISCUSSION

This paper presents a new model that represents user message exchanges and their influences in providing information to each other. We capture information seeking and information providing roles of users as well as message roles within the discussion, and combine them with global influences of the participants. The resulting model is used in identifying the most useful messages in answering the initial questions. We plan to use the identified best answers in helping other students with similar issues. The RiNet model can be also used in summarizing student contributions for teachers, such as in reporting significant seekers or providers.

We applied 5 different parameters $\omega$ and $\gamma$ parameter combination setting to show different MRR score however did not investigate parameter-searching methods. To find the best parameters, we can employ cross validation; use training set to search for the best performance parameters using MRR score and apply it to test set. One simple parameter selection method is grid points search. Since there are only two parameters, the number of search points is not too large. Furthermore grid search can be easily parallelized for distributed computing. Other parameter-searching methods such as conjugate gradient-descent, genetic algorithm, simulated annealing, or stochastic search can be applied with our objective function, ranking based MRR score. In the future, we will present an efficient procedure to estimate parameters of the model.

For a deeper analysis of message influence in discussions, we need to take into account discussion topics. Discussion contributions can be clustered based on their focus or question topics, which can help us find useful messages per topic clusters. We plan to apply the RiNet model to other Q&A forums, as we expect similar information roles.

ACKNOWLEDGEMENT

This work is supported by the National Science Foundation, CISE IIS (award #0917328) and REEESE (award #1008747) programs.